\documentstyle[preprint,floats,tighten,aps]{revtex}

\begin{document}

\draft

\title{
Extension of Quantum Molecular Dynamics and \\
its Application to Heavy-Ion Collisions}

\author{
Toshiki Maruyama$^1$,
Koji Niita$^{1,2}$ and Akira Iwamoto$^1$}

\address{
1 Advanced Science Research Center,\\
Japan Atomic Energy Research Institute, \\
Tokai, Ibaraki, 319-11 Japan}
\address{
2 Research  Organization for Information Science \& Technology, \\
Tokai, Ibaraki, 319-11 Japan}

\maketitle

\begin{abstract}
In order to treat low-energy heavy-ion reactions,
we make an extension of quantum molecular dynamics method.
A phenomenological Pauli potential is introduced into
effective interactions to approximate the nature of the
Fermion many-body system.
We treat the widths of nucleon wave-packets as
time-dependent dynamical variables.
With these modifications, our model can well describe the
ground-state properties in wide mass range.
Improvements due to the extension are also obtained in the
nucleus-nucleus collision calculations.
\end{abstract}


\section{Introduction}

With the progress of computers, microscopic simulation methods
have become popular in the heavy-ion reaction studies.
The benefit of microscopic simulation method is
that one can investigate nuclear reactions without making
any specific assumption on the reaction mechanism.
There are many kinds of microscopic simulations such as
the time-dependent Hartree Fock (TDHF) \cite{refTDHF}
which is a mean-field theory,
Vlasov equation which is the semiclassical approximation of TDHF,
Vlasov-Uehling-Uhlenbeck (VUU) equation or
Boltzmann-Uehling-Uhlenbeck (BUU) equation \cite{refVUU}
in the other name,
which consist of Vlasov and the two-body collision term,
the Cascade model \cite{refCascade} which includes only
two-body collision term, and so on.
Especially, VUU/BUU equation, which includes both of
mean-field and two-body collision term,
has become a standard framework for the heavy-ion reaction study
from the low or intermediate energy to the high energy region.
However, VUU/BUU equation, which is basically one-body theory,
has a difficulty in dealing with the phenomena of fluctuation
such as the fragment formation.

Molecular dynamics approaches like quantum molecular dynamics
(QMD) \cite{refQMD,refQMDrep}
have been developed in order to calculate the fragmentation process.
In QMD, we assume a single-particle distribution function of
a nucleon as a Gaussian wave packet, and calculate the time evolution
of the system according to the classical Newtonian equation of motion
plus the two-body collision term.
It contains both aspects of the mean-field and the two-body collisions,
and is applicable to the wide energy-region.
Furthermore, QMD can deal with the fragment formation since it is
based on a many-body framework which traces the motion of each nucleon.

Due to its applicability and its ability to describe
fragment formation, QMD is widely used recently for heavy-ion
reactions from intermediate to high energies.
For low-energy reactions such as fusion, fission and deep
inelastic collision process, however, microscopic simulations
by using the molecular dynamics have not been
studied except for a few works.
One example is the analysis of the fusion reaction
using QMD \cite{refQMDfusion}.
It was reported, however, that several extra
nucleons are emitted during the collision process in the calculation.
This is due to the insufficient stability of initial ground state nuclei.
We have to settle this problem to study low-energy collisions
of heavy-ions  using the molecular dynamics.

For the simulation of low-energy nuclear reactions,
frameworks with anti-symmetrization of the total wave function
like FMD \cite{refFMD} and AMD \cite{refAMD} are
suitable for describing ground state properties and
reaction processes.
However, they are not applicable for very heavy systems
since they need much CPU time which is approximately
proportional to the fourth power of the particle number.
For calculations of heavy systems, application of the QMD
framework (without anti-symmetrization) is still necessary.

In this paper we propose an extended version of QMD method
in view of the simulation of low-energy phenomena.
Its applicability to treat ground state properties and
nuclear reaction is investigated.
Several improvements over standard QMD are found in this study.
In Sec.~2 we describe the formalism of the extended QMD method,
and the ground state properties by our model is discussed in Sec.~3.
Its application to the nuclear reaction is reported in Sec.~4.
Finally, summary and the discussion is given in section 5.

\section{Extension of QMD}

The insufficient stability of the standard QMD ground state is mainly
due to the fact that they are not at their energy-minimum states.
In the standard QMD, the kinetic energy term arising
from the momentum variance of wave-packets is spurious
and we do not take into account this term.
Thus the constituent nucleons have finite momenta and are moving
around in the ground state with appropriate binding energies.
If we take the energy-minimum states, all the nucleons stop
their motion and get into the over-bound states where
the Pauli principle is broken.
To solve this difficulty we make an extension
of QMD in two points so as to take energy-minimum
state as an initial ground nucleus:
First, we include the so-called Pauli potential into
effective interactions \cite{refOhnishi} in order to
approximate the nature of Fermion many-body system.
Second, we take into account the kinetic-energy term of
the momentum variance of wave-packets to the Hamiltonian.
In accordance with this we treat the width of each wave packet
as a dynamical variable \cite{refValta}.
We call here this extension of QMD as ``EQMD''.

\subsection{Equation of motion of the system}

The equation of motion is obtained by the time-dependent
variational principle.
We assume the total wave function of the system
as a direct product of Gaussian wave packets of nucleons
\begin{eqnarray}
\Psi &=& \prod_i  \phi_i({\bf r}_i)\ ,                          \\
\phi_i({\bf r}_i)&=&\left({\nu_i+\nu_i^*\over 2\pi}\right)^{3/4}
   \exp\left[-{\nu_i\over2}({\bf r}_i-{\bf R}_i)^2
             +{i\over \hbar}{\bf P}_i\cdot{\bf r}_i\right]\ .
\end{eqnarray}
Here ${\bf R}_i$ and ${\bf P}_i$ are the centers of
position and momentum of the $i$-th wave packet.
We introduce the complex Gaussian width $\nu_i$ as
\begin{equation}
\nu_i\equiv{1\over\lambda_i}+i\delta_i ,
\end{equation}
where $\lambda_i$ and $\delta_i$ are its real and  imaginary parts,
respectively.
The Hamiltonian is written as
\begin{eqnarray}
H
&=&\langle\Psi|\sum_i-{\hbar^2\over 2m}\nabla_i^2
 -\hat T_{\rm CM} +\hat H_{\rm int} |\Psi\rangle \\
&=&\sum_i\left[{{\bf P}_i^2\over 2m}
             +{3\hbar^2(1+\lambda_i^2\delta_i^2)\over 4m\lambda_i}\right]
	 -T_{\rm CM} +H_{\rm int} \ ,
\end{eqnarray}
where $T_{\rm CM}$ and $H_{\rm int}$ denote the spurious
zero-point center-of-mass kinetic energy and the potential
energy term, respectively.
Here, one should note that the Hamiltonian includes terms
originating from momentum variances of wave packets which were
neglected as spurious constant terms in the standard QMD.
Subtraction of the spurious zero-point CM kinetic energy
is necessary since we include the kinetic energy from
the momentum variance of wave-packets.

The equations of motion of these variables
are determined by the time-dependent variational principle
\begin{eqnarray}
&&\delta \int_{t_1}^{t_2} {\cal L} {\rm d} t=0\ ,           \\
&&{\cal L} (\{{\bf R}_i,{\bf P}_i,\lambda_i,\delta_i,
           \dot{\bf R}_i,\dot{\bf P}_i,\dot \lambda_i,\dot \delta_i\})
\equiv \langle\Psi|i\hbar{d \over dt}-\hat H|\Psi\rangle .
\end{eqnarray}
Then we get 8$A$ dimensional classical equations of motion, i.e.
equations of motion of 4$A$ parameters
($A$ means the number of constituent particles)
\begin{eqnarray}
&&\dot{{\bf R}_i}={\partial H\over\partial{\bf P}_i}\ ,
\ \ \
\dot{{\bf P}_i}=-{\partial H\over\partial{\bf R}_i}\ ,  \nonumber\\\\
&&{3\hbar\over4}\dot\lambda_i=-{\partial H\over\partial\delta_i}\ ,
\ \ \
{3\hbar\over4}\dot\delta_i={\partial H\over\partial\lambda_i}\ . \nonumber
\end{eqnarray}

\subsection{Subtraction of zero-point CM kinetic energy}

We subtract the zero-point center-of-mass kinetic energy
of the system $T_{\rm CM}$ following the basic idea of Ref.~\cite{refAMD}.
In our case, however, all the wave-packets have different
contributions to zero-point kinetic energy.
Therefore we take $T_{\rm CM}$ as
\begin{equation}
T_{\rm\scriptscriptstyle CM}
  =\sum_i {t^{\rm\scriptscriptstyle CM}_i\over M_i}\ ,
\end{equation}
where
$t^{\rm\scriptscriptstyle CM}_i$ is the zero-point kinetic energy
of the wave-packet $i$ written as
\begin{equation}
t^{\rm\scriptscriptstyle CM}_i = {\langle\phi_i|\nabla^2|\phi_i\rangle\over2m}
                  -{\langle\phi_i|\nabla|\phi_i\rangle^2\over2m}\ .
\end{equation}
and $M_i$ is the ``mass number'' of the fragment
to which the wave-packet $i$ belongs.
The ``mass number'' is calculated as the sum
of the ``friendships'' of other nucleons
\begin{eqnarray}
M_i&=& \sum_j F_{ij}  \\
F_{ij}&\equiv&
\cases{
\hbox to 4cm{1}
    (|{\bf R}_i-{\bf R}_j|<a)\cr
\hbox to 4cm{$e^{-(|{\bf R}_i-{\bf R}_j|-a)^2/b}$}
    (|{\bf R}_i-{\bf R}_j|\geq a)
}\ ,
\end{eqnarray}
where we use the parameters $a=1.7\ \rm fm$ and $b=4\ {\rm fm}^2$.

\subsection{Effective interaction}

For the effective interaction, we use
Skyrme,  Coulomb, Symmetry
and the Pauli potential
\begin{equation}
H_{\rm int}=H_{\rm Skyrme}+H_{\rm Coulomb}
	   +H_{\rm Symmetry}+H_{\rm Pauli}\ .
\end{equation}


For the form of Skyrme interaction,
we use the most simple one
\begin{eqnarray}
H_{\rm Skyrme}
&=&{\alpha\over2\rho_0} \int\rho^2({\bf r}){\rm d}^3r
 +{\beta\over(\gamma+1)\rho_0^{\gamma}}
      \int\rho^{\gamma+1}({\bf r}) {\rm d}^3r      \\
\rho({\bf r})&=&\sum_i^A\rho_i({\bf r}) \ ,     \\
\rho_i({\bf r})&=&{1\over(\pi\lambda_i)^{3/2}}
     \exp[-({\bf r}-{\bf R}_i)^2/\lambda_i] \ .
\end{eqnarray}
In the treatment of real system, however, we exclude the
self interaction
\begin{eqnarray}
H_{\rm Skyrme}
=&&{\alpha\over2\rho_0} \sum_{i,j\neq i}
\int \delta({\bf r}_i-{\bf r}_j)
\rho_i({\bf r}_i)\rho_j({\bf r}_j){\rm d}^3r_i{\rm d}^3r_j   \nonumber\\
&&+ {\beta\over(\gamma+1)\rho_0^{\gamma}} \sum_{i,j\neq i}
\int \delta({\bf r}_i-{\bf r}_j)\rho({\bf r}_i)^{\gamma-1}
\rho_i({\bf r}_i)\rho_j({\bf r}_j){\rm d}^3r_i{\rm d}^3r_j\ ,      \\
\equiv&& H_2+H_{\gamma+1} \ .
\end{eqnarray}
For the numerical calculation of the density-dependent
term $H_{\gamma+1}$,
we perform a three-fold loop computation since we use the
density-dependent term with $\gamma=2$ which is identical
to the three-body interaction.
Although the approximate treatment of density-dependent term
with the two-fold loop computation much reduces the CPU time,
it causes much ambiguity.


We also employ the symmetry potential as
\begin{equation}
H_{\rm Symmetry}={c_{\scriptscriptstyle\rm S}\over 2\rho_0}\sum_{i,j\ne i}
\int[2\delta(T_i,T_j)-1]\rho_i({\bf r})\rho_j({\bf r}){\rm d}^3r\ ,
\end{equation}
where $T_i$ is the isospin index of nucleon $i$.
In the nuclear matter limit, this term goes to
\begin{equation}
\int{c_{\scriptscriptstyle\rm S}\over 2}
  {(\rho_{\rm p}-\rho_{\rm n})^2\over\rho_0}{\rm d}^3r\ ,
\end{equation}
where $\rho_{\rm p}$ and $\rho_{\rm n}$ denote
proton and neutron densities, respectively.
%

\subsection{Pauli potential}
We introduce a phenomenological repulsive potential
which inhibits nucleons of the same spin $S$ and isospin $T$
to come close to each other in the phase space.
We assume a very simple form for this potential,
\begin{eqnarray}
H_{\rm Pauli}&=&{c_{\rm\scriptscriptstyle P}\over 2}\sum_i(f_i-f_0)^\mu
\theta(f_i-f_0)\ ,
                                                             \\
f_i&\equiv&\sum_j  \delta(S_i,S_j)\delta(T_i,T_j)
                  \left|\langle\phi_i|\phi_j\rangle\right|^2\ ,
\end{eqnarray}
where $f_i$ is the overlap of a nucleon $i$ with the
same kind of nucleons (including itself),
and we take the threshold parameter $f_0 \approx 1$.
$c_{\rm\scriptscriptstyle P}$ is the strength of the potential.


In order to see the statistical behavior of infinite system with
our Pauli potential, we show in Fig.~1 the energy per nucleon of
free nucleon gas with the density- and temperature-dependence.
In the limit of infinite system, we assume all the wave packets
approach to plane waves with uniform coordinate-space distribution.
Then we perform the Metropolice simulation in the momentum space
as in Refs.~\cite{refPauliA,refPauliB}.
As for the parameters of Pauli potential in the figure,
we take $c_{\rm\scriptscriptstyle P}=15$~MeV, $f_0=1.05$ and $\mu=2.0$.
No nuclear potential and Coulomb potential is included.
In the low-temperature region our gas is closer to
the Fermi gas rather than the Boltzmann gas.
In the high-temperature region, however, it approaches
to the classical limit of the Boltzmann gas rapidly.
Although this is a general tendency with any set of parameters,
the behavior of our gas is apparently different from
that of the simple classical gas and we can describe,
to a certain extent, the property of Fermi gas.
For the ground state nuclei and for low- and medium energy
reactions treated in this paper, the inclusion
of Pauli potential greatly improves the standard QMD
as seen in this paper.

\begin{figure}
\vspace{170mm}\hspace{-5mm}
\includegraphics{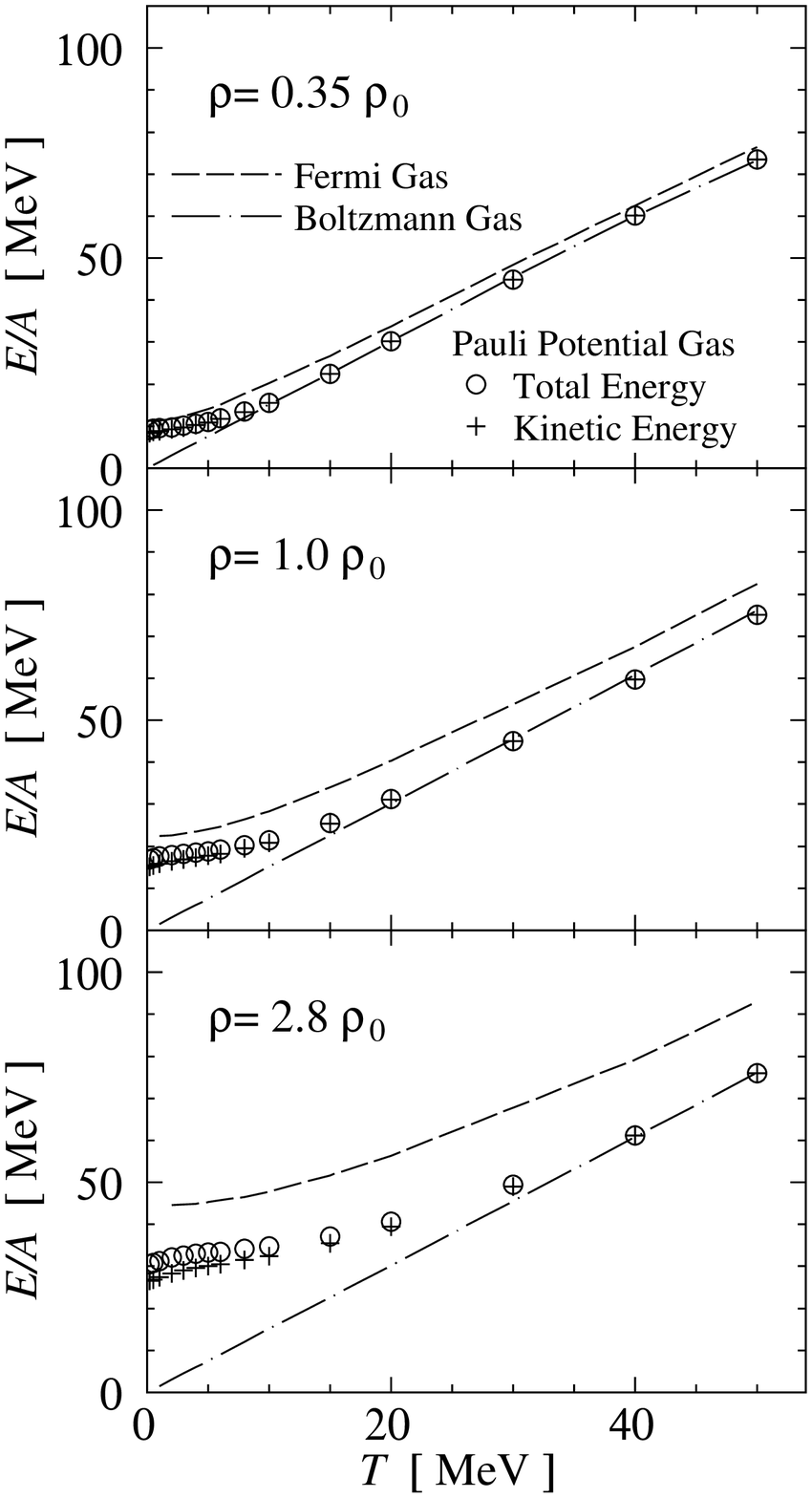}
\vspace{0mm}\hspace{90mm}
\caption{
Energies per nucleon of infinite system without nuclear
and Coulomb potentials.
Open circles and crosses denote the total and the kinetic
energy of the gas with Pauli potential with
$c_{\rm\scriptscriptstyle P}=15$~MeV,
$f_0=1.05$ and $\mu=2.0$.
Dashed and dot-dashed lines denote Fermi gas and the
classic Boltzmann gas.}
\end{figure}

There are two possible ways to fix the interaction parameters.
One is to keep the saturation condition of nuclear matter
with nuclear and Pauli potentials.
In this case we have to adjust the Skyrme interaction parameters
as well as the Pauli potential.
We calculate the energy of the infinite system at
zero-temperature as a function of its density.
Then the Skyrme interaction parameter is adjusted to give
16 MeV binding energy at the saturating point.
Keeping the saturation property of nuclear matter,
we search good parameters of Pauli potential for
finite nuclear systems.
We call the parameters fixed in this way as ``parameter set 1''.
Another way is to fix the Skyrme parameters in their
original values neglecting the contribution of Pauli
potential to the saturation property of matter.
In this case we adopt standard parameters of the Skyrme interaction.
The parameters of Pauli potential is searched to give a good agreement
to the  systematic trend of binding energies of finite systems.
We call these parameters ``parameter set 2''.
The values of the parameter of both sets are listed in Table I.

\begin{table}
\caption{
Parameter values used in the Skyrme, Symmetry and Pauli potentials.
}
\begin{tabular}{ccccccccc}
&                 & $\alpha$     & $\beta$    & $\gamma$
& $c_{\scriptscriptstyle\rm S}$
   & $c_{\rm\scriptscriptstyle P}$ & $f_0$ & $\mu$ \\
\noalign{\hrule}
& parameter set 1 & $-116.6$ MeV & $70.8$ MeV &   2  & 25 MeV
   &   15 MeV       & 1.05  & 2.0   \\
& parameter set 2 & $-124.3$ MeV & $70.5$ MeV &   2  & 25 MeV
   &   15 MeV       & 1.0   & 1.3   \\
\end{tabular}
\end{table}

\subsection{Two-body collision term}

For the treatment of two-body collisions,
we follow the prescription of the standard QMD.
If a pair of two nucleons fulfill these conditions, i.e.,
i) their relative distance takes its minimum value within the
time-step, and ii) the minimum distance
being smaller than a certain value $d_{\rm coll}$,
then a stochastic two-body collision is set to occur iii) with
the probability $P_{\rm coll}$ decided as
\begin{equation}
P_{\rm coll}={\sigma_{\rm NN}\over \pi {d_{\rm coll}}^2}\ .
\end{equation}
Here we take 2.0 fm for the value of $d_{\rm coll}$
and $\sigma_{\rm NN}$ is the energy-dependent
nucleon-nucleon collision cross section parameterized \cite{refAMD}
given as
\begin{eqnarray}
\sigma_{\rm NN}
&=&{100\over 1+\epsilon/ 200\ {\rm MeV}}\ {\rm mb}\ ,
\\
\epsilon&\equiv& p_{\rm rel}^2/2m\ .
\end{eqnarray}
In this paper the angular distribution of the final
state of collision is assumed to be isotropic.
The final relative momentum of colliding nucleons is
searched so as to conserve the total energy of the system,
since the total energy conservation is not guaranteed due to the
momentum dependence of Pauli potential.
The Pauli principal is also checked and the collisions
with unsatisfactory final states are canceled.

\section{The ground state properties}

One has to prepare energy-minimum states as initial ground nuclei.
They are obtained by starting from a random configuration
and by solving the damped equations of motion as
\begin{eqnarray}
&&\dot{{\bf R}_i}=               {\partial H\over\partial{\bf P}_i}
               +\mu_{_{\bf R}}{\partial H\over\partial{\bf R}_i} \ ,
\ \ \
\dot{{\bf P}_i}=              -{\partial H\over\partial{\bf R}_i}
         +\mu_{_{\bf P}}{\partial H\over\partial{\bf P}_i}\ ,  \nonumber\\
      \label{eqDamp}\\
&&{3\hbar\over4}\dot\lambda_i=            -{\partial H\over\partial\delta_i}
          +\mu_{_{\lambda}}{\partial H\over\partial\lambda_i}\ ,
\ \ \
{3\hbar\over4}\dot\delta_i=               {\partial H\over\partial\lambda_i}
                    +\mu_{_{\delta}}{\partial H\over\partial\delta_i}\ .
\nonumber
\end{eqnarray}
Here $\mu_{_{\bf R}}$, $\mu_{_{\bf P}}$, $\mu_{_{\lambda}}$
and $\mu_{_{\delta}}$ are damping coefficients.
With negative values of these coefficients the system
goes to its (local) minimum point,
\begin{eqnarray}
{dH\over dt}&=&\sum_i\left[
{\partial H\over\partial{\bf R}_i}\dot{\bf R}_i+
{\partial H\over\partial{\bf P}_i}\dot{\bf P}_i+
{\partial H\over\partial\lambda_i}\dot\lambda_i+
{\partial H\over\partial\delta_i}\dot\delta_i\right]             \\
&=&\sum_i\left[
\mu_{_{\bf R}}\left({\partial H\over\partial{\bf R}_i}\right)^2+
\mu_{_{\bf P}}\left({\partial H\over\partial{\bf P}_i}\right)^2+
{4\mu_{_{\lambda}}\over3\hbar}\left({\partial H\over\partial\lambda_i}
\right)^2+
{4\mu_{_{\delta}}\over3\hbar}\left({\partial H\over\partial\delta_i}
\right)^2\right].
                                                         \\
&\leq 0
\end{eqnarray}
All the nucleon wave-packets obeying Eq.~\ref{eqDamp} stop their motions
at the energy-minimum state,
\begin{eqnarray}
\dot{\bf R}_i&=&0\ , \ \ \ \ \
\dot{\bf P}_i=0\ , \nonumber\\
\\
\dot\lambda_i&=&0\ , \ \ \ \ \
\dot\delta_i=0\ , \nonumber
\end{eqnarray}
while in the standard QMD model nucleons are
moving around in the ground-states.
Fermi motions are shared by the momentum variance of wave-packets
and the non-zero values of momenta ${\bf P}_i$ which come from the
momentum-dependence of Pauli potential.

\subsection{Mass number dependence of the binding energy}

The binding energy of a nucleus in the present framework is
obtained from the minimum-energy condition for the ground state,
while in the standard QMD framework it is only an input to
fit the empirical value.
Our interest is therefore how well the calculated values
reproduce the experimental data.
Figure 2 shows the comparison of binding energy per nucleon
$E_{\rm bind}$ of our calculation and the experimental value.

\begin{figure}
\vspace{-15mm} \hspace {10mm}
\includegraphics{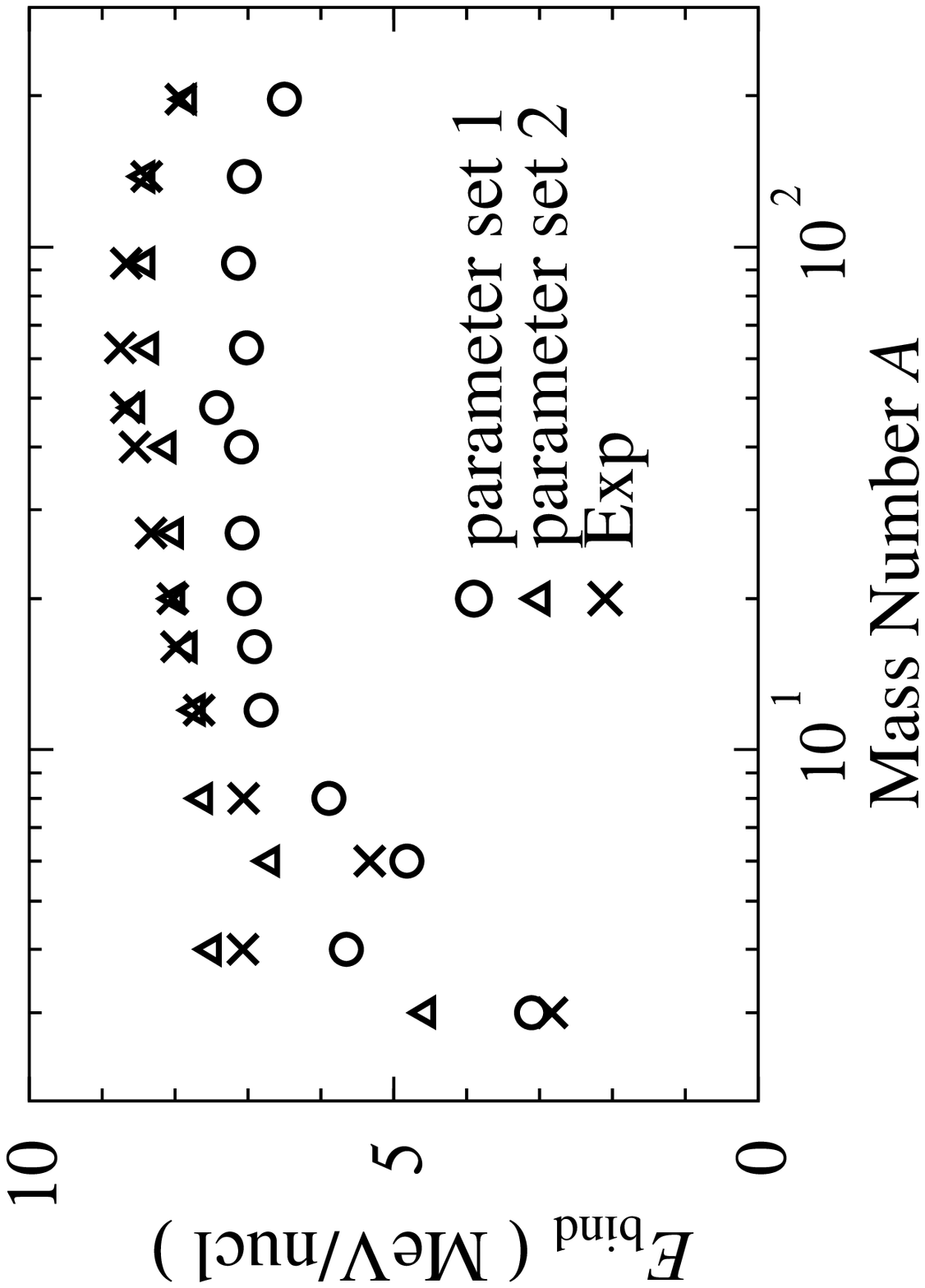}
\vspace{100mm}
\caption{
Binding energies per nucleon of ground state nuclei.
Circles and triangles denote our model with
parameter set 1 and 2, crosses denote
corresponding experimental values.
}
\end{figure}

With interaction parameter set 1 which guarantees the infinite
matter properties, we can reasonably reproduce the binding
energies of finite systems shown with circles in Fig.~2.
With parameter set 2 (shown with triangles in Fig.~2)
which uses the standard value of Skyrme interaction
without considering the matter properties,
we can reproduce the binding energy of nucleus almost perfectly.

\subsection{Some typical features of individual nuclei}

The shapes and density distributions of ground states
are also well reproduced.
In the case of light nuclei, e.g., $^{12}{\rm C}$, we can reproduce
the alpha-clustering structures as displayed in Fig.~3.
On the left are shown the radial density distribution (upper)
and distribution of the real parts of nucleon wave-packet widths (lower)
and on the right is shown the contour plot of density.
This $^{12}{\rm C}$ nucleus is calculated with parameter set 2 and
has the binding energy of 92.7 MeV
and the rms radius of 2.31 fm which agree to the
experimental values (92.2 MeV and 2.46 fm \cite{refExpCarbonOxigen}).
By using interaction parameter set 1, we can also make a very similar
ground state of $^{12}{\rm C}$ except for the binding energy.
This three-alpha structure of $^{12}{\rm C}$ is almost the same as
those obtained in AMD \cite{refAMD} and FMD \cite{refFMD}.
This gives a support of introducing the Pauli potential as
a phenomenological substitute of anti-symmetrization.

For heavy nuclei, the density profiles is shown
in the upper parts of Fig.~4.
We see that the general feature of the density and the
surface thickness are well reproduced in our calculations.
The value of density, however, is somewhat higher than normal matter density.
This is one of the problems of our model which makes the size of nuclei
and therefore the reaction cross section slightly smaller than the data.
We guess one reason is the Pauli principle is not perfectly
realized in the ground nuclei since the Pauli potential
we are using is rather moderate.
One can see this fact in the Fig.~1 at low temperature
where our system is slightly deviating from the Fermi gas.

The lower parts of Fig.~4 shows the distribution of
the real parts of nucleon wave-packet widths.
The abscissas represent the distance of
the wave packet from the center.
Wave packets near the center are spatially
more spread than those near the surface.
Note that this fact does not mean the kinetic-energy density
is higher at the surface since the density and the imaginary
part of the width also contribute to the kinetic-energy density.
In fact we obtain the kinetic-energy density distribution
higher at the center and lower at the surface of a nucleus.
A narrow width for a surface nucleon is required because
otherwise the surface diffuseness becomes unrealistically large.
A wide width for a central nucleon is also expected because
our model predicts naturally a kind of matter limit
(large width limit) for the nucleon sitting in the
central part of the nucleus of large mass limit.

\begin{figure}
\vspace{-15mm}\hspace{2mm}
\includegraphics{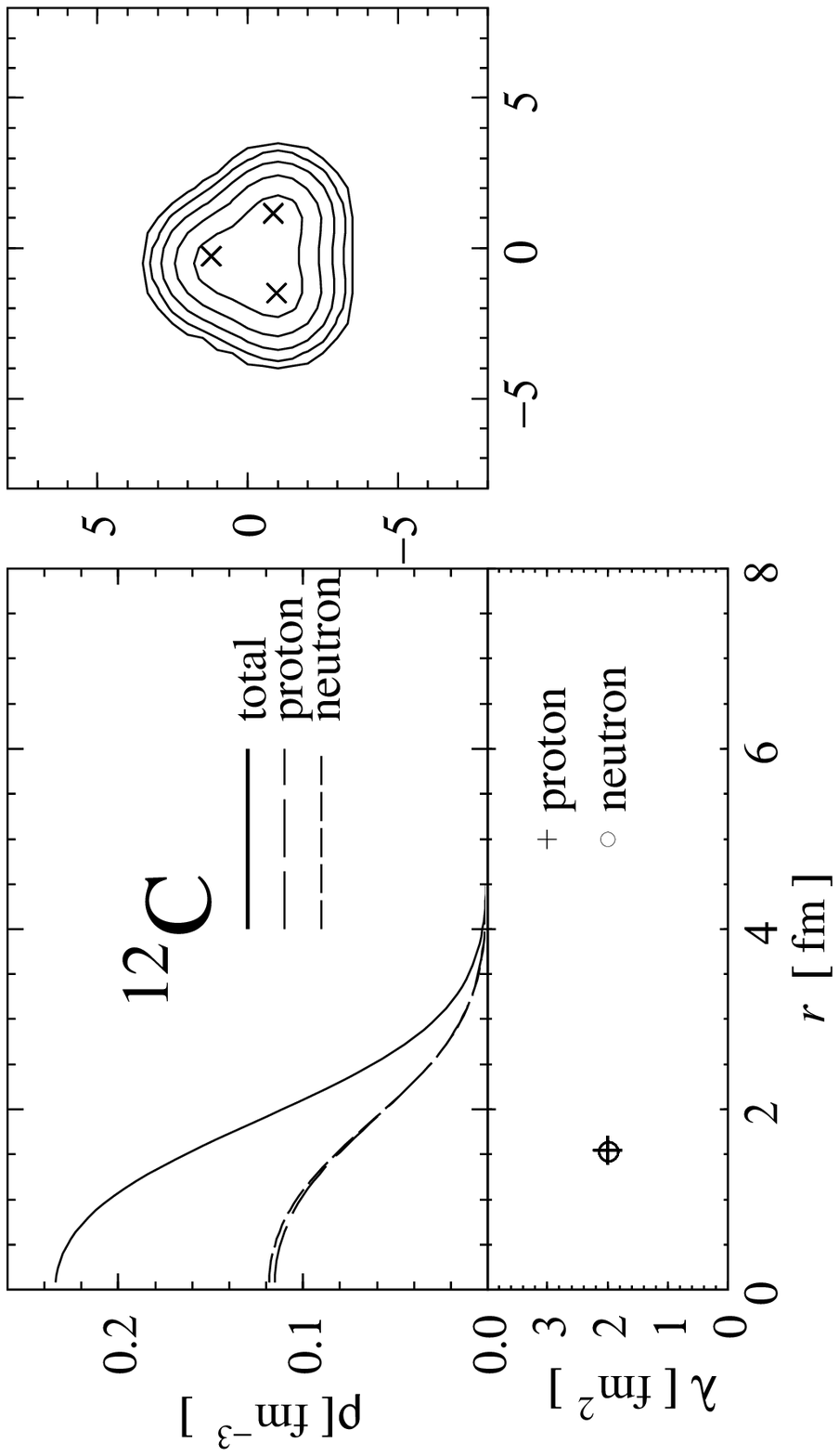}
\vspace{80mm}
\caption{
Radial distribution (upper-left),
the contour plot (upper-right) of density and
the real part of the Gaussian widths (lower part) of
$^{12}{\rm C}$ ground state.
Interaction parameter set 2 is used.
}
\end{figure}
\begin{figure}
\vspace{-15mm}\hspace{-8mm}
\includegraphics{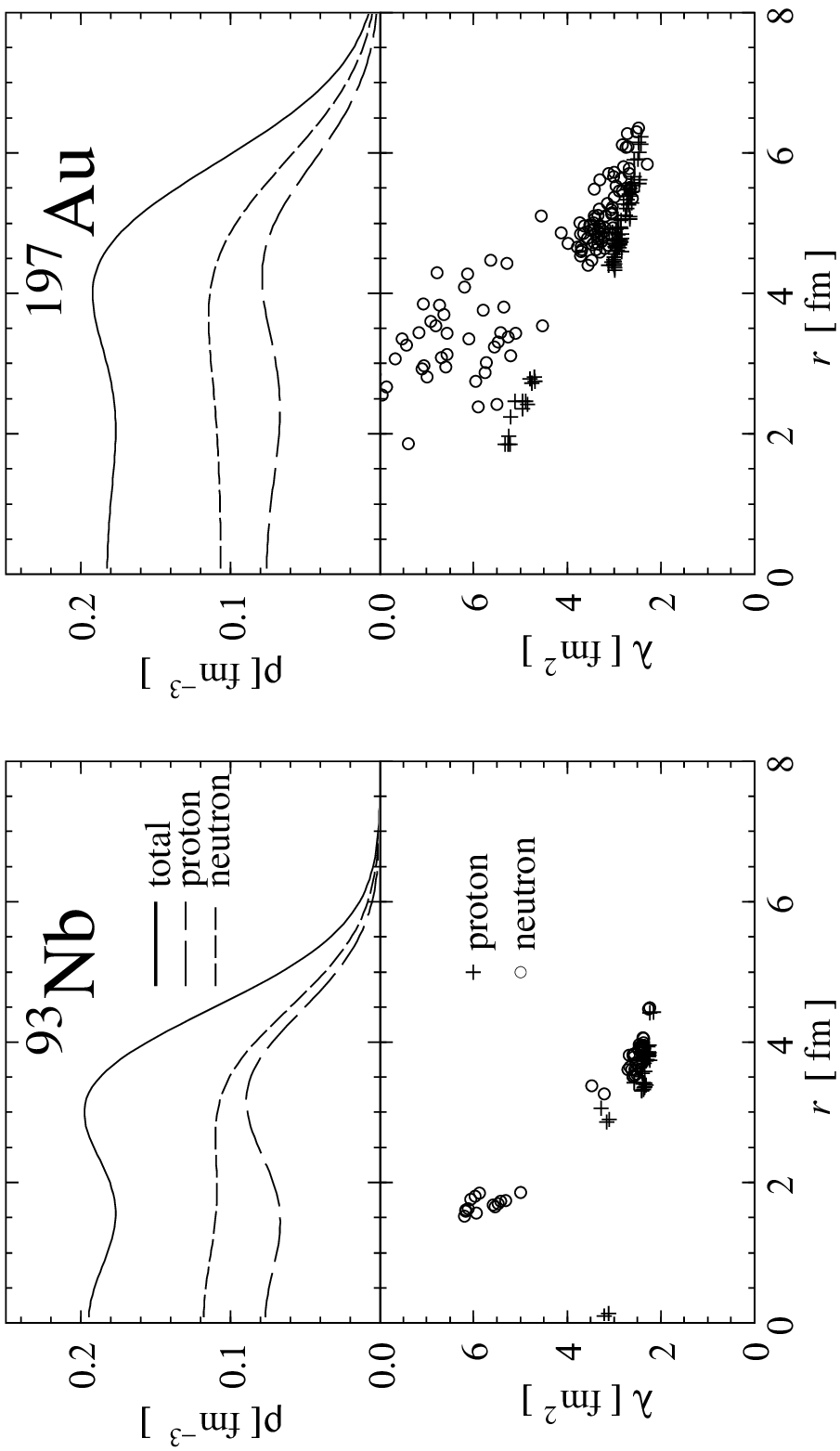}
\vspace{95mm}
\caption {
Density distribution (upper parts) and Gaussian widths
(lower parts) of the ground states of $^{93}{\rm Nb}$ and $^{197}{\rm Au}$.
Abscissas for both denote the distance from the center.
Interaction parameter set 2 is used.
}
\end{figure}

\section{Nucleus-nucleus collisions}

We apply our model to nucleus-nucleus collisions in this section.
In the calculation of nuclear reactions we boost two initial nuclei
according to the incident energy and then solve the EQMD equation of
motion together with the two-body collision term as in the standard QMD.
For all the fragments produced, their statistical decay processes to
the final products are calculated \cite{refQMDplusStat}.
Typically, within a time-scale of about 100 fm/c the dynamical
part of nuclear reaction is completed, and is followed by
the statistical decay process for a time-scale of
several-order longer \cite{refQMDplusStat,refQMDrel}.
Since it is not practical nor reliable to treat this statistical
process in a framework of simulation, we adopt
the standard statistical model \cite{refQMDplusStat,refCascade}
for the latter process.
With this hybrid model of QMD plus statistical decay calculation,
we can calculate observables such as the mass distribution
and the energy spectra of fragments.

\subsection{Fragment mass distribution in the medium energy collisions}

Figure 5 shows our calculation of fragment production cross sections in
the $^{12}{\rm C}+^{12}{\rm C}$ (29 MeV/nucleon) reaction compared with
the standard QMD (QMDstd) and AMD.
The AMD results \cite{refAMD} reproduce the experimental
data well especially for light fragments.
Solid lines show the final fragment distribution after the statistical
decay calculation while dashed lines show fragments
produced in the dynamical process before the statistical decay.

Though the final results of three models are quite similar,
there are some differences between them  before the statistical decay.
Especially, AMD and QMDstd apparently differ with each other:
In the AMD result, enhancement of $A_{\rm f}$ = 4 and 8
($N$ alpha fragments) is seen while
there is no peak at 4$N$ in the QMDstd result.
This is mainly because AMD can describe three-alpha
structure of $^{12}{\rm C}$ while QMDstd can not.
There exist peaks at $A_{\rm f}$=4 in the
present results of EQMD with parameter sets 1 and 2.
Dynamical emissions of alpha clusters are enhanced
due to the improvement of ground states in our model.

To see the effect of dynamical treatment of wave-packet width,
we compare in Fig.~6 the full EQMD calculation with that of
fixed-width constraint for the same quantity as Fig.~5.
In the fixed width calculation we solve the equation of motion
only for ${\bf R}_i$ and ${\bf P}_i$ keeping the widths
$\nu_i \equiv \lambda_i^{-1}+i\delta_i$ as constants
(6$A$-dimensional calculation) using exactly the same
interactions and the initial conditions as in the full EQMD.
Thus this calculation also differs from the QMDstd.
With fixed wave-packet widths, the distribution of dynamically
produced fragments has strong peaks at $A_{\rm f}=4N$ and
productions of other fragments and nucleons are rather hindered.
Underestimation of nucleon yield is seen even after the statistical decay.
For the description of nucleon emission process,
the dynamical treatment of wave-packet widths is
essential in the QMD calculation.

\begin{figure}
\vspace{170mm}\hspace{-3mm}
\includegraphics{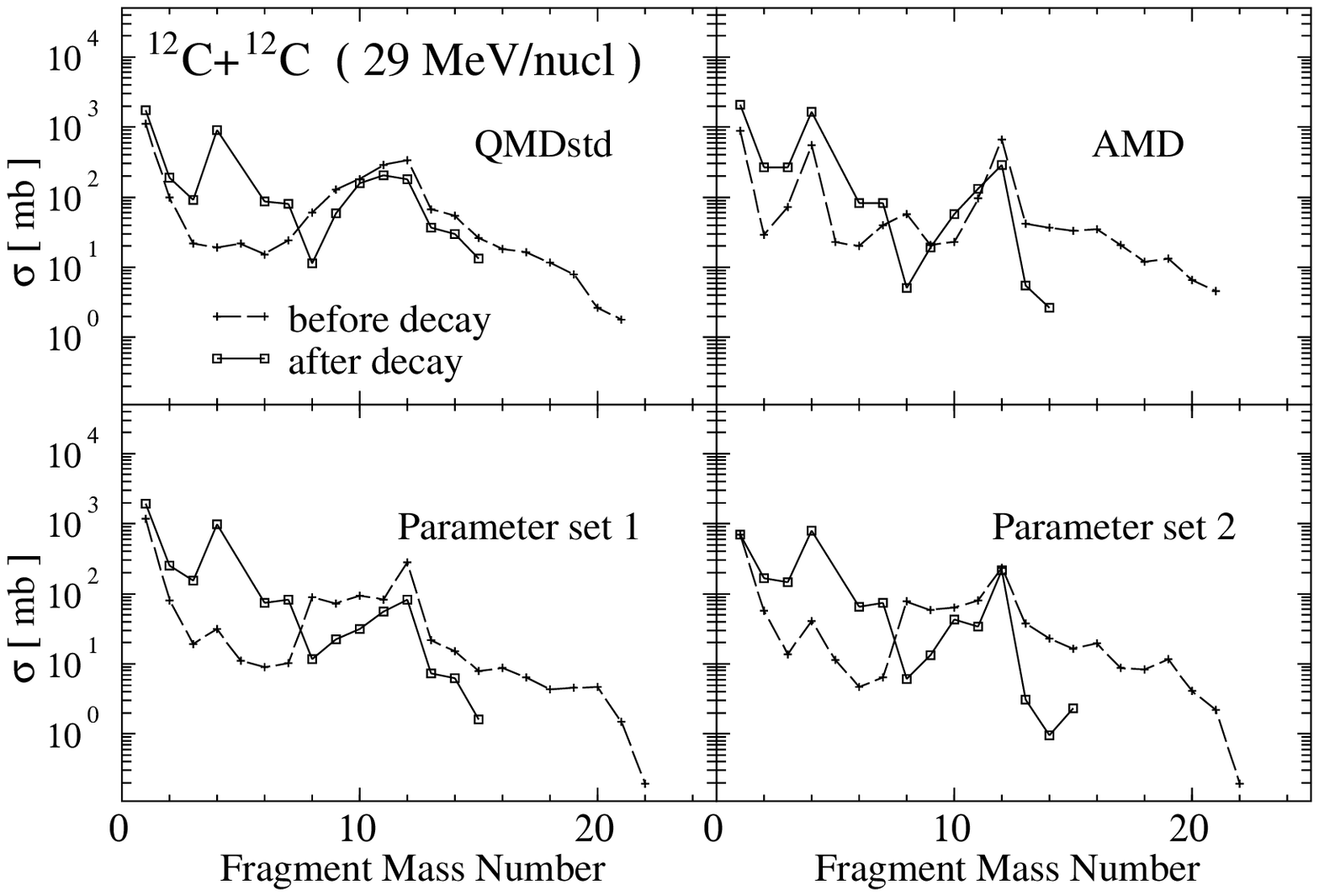}
\vspace{-75mm}
\caption {
Fragment mass distribution in
$^{12}{\rm C}+^{12}{\rm C}$ (29 MeV/nucleon) reaction
calculated with standard QMD, AMD and EQMD with parameter sets 1 and 2.
Dashed lines denote fragments produced in dynamical process
before statistical decay calculation,
while solid lines denote final fragments after statistical decay.
}
\end{figure}

\begin{figure}
\vspace{165mm}\hspace{-2mm}
\includegraphics{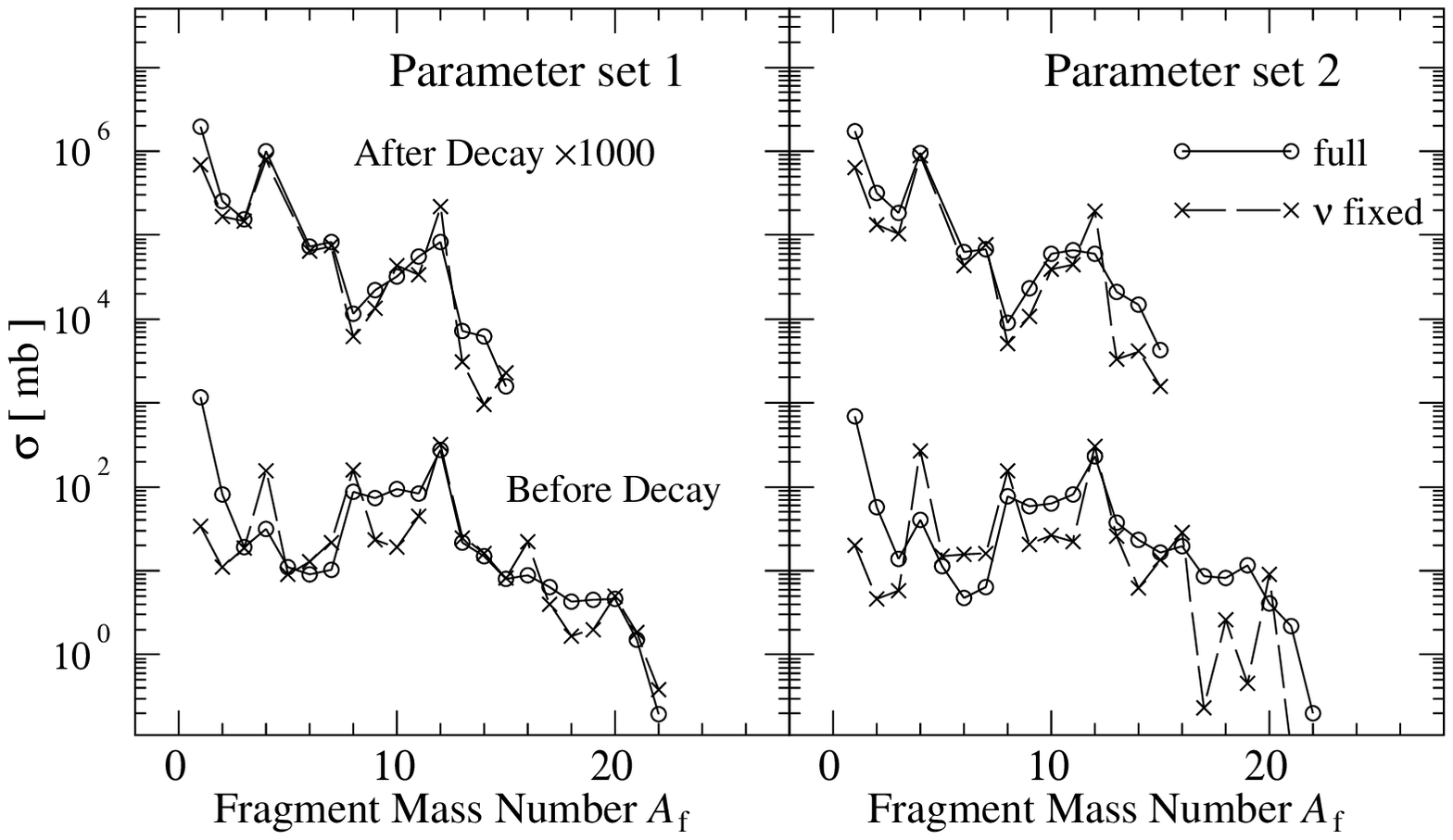}
\vspace{-80mm}
\caption{
Fragment mass distribution in
$^{12}{\rm C}+^{12}{\rm C}$ (29 MeV/nucleon) reaction.
Dashed lines denote fragments produced in dynamical process
before statistical decay calculation,
while solid lines denote final fragments after statistical decay.
}
\end{figure}

\subsection{Fusion cross section of $^{16}{\rm O}$+$^{16}{\rm O}$ reaction}

We examine our model in the case of the
$^{16}{\rm O}+^{16}{\rm O}$ fusion reaction.
The same reaction has been calculated by the
standard QMD model in Ref.~\cite{refQMDfusion}.
It was found there that if one admits events of up to
three nucleons escape into the fusion events, fusion cross
section is nicely reproduced by the standard QMD model.
If one demands no nucleon emission, however,
the calculated fusion cross section is almost zero.
In the present calculation, we demand that in the fusion event
there is no nucleon/fragment emission within a certain time
(in this paper we have chosen 450 fm/c after the contact of two nuclei).
Classification of nucleons into fragments is done by the condition
that nucleons within the distance of 3.0 fm belong to the same fragment.
Since we judge the fusion events at finite time,
we don't calculate the statistical decay process in this case.
This calculation corresponds to a complete fusion reaction,
and such exclusive calculation is possible in EQMD owing to
the fact that EQMD ground state is stable enough so that no
spurious particle emission happens.

\begin{figure}
\vspace{-15mm}\hspace{10mm}
\includegraphics{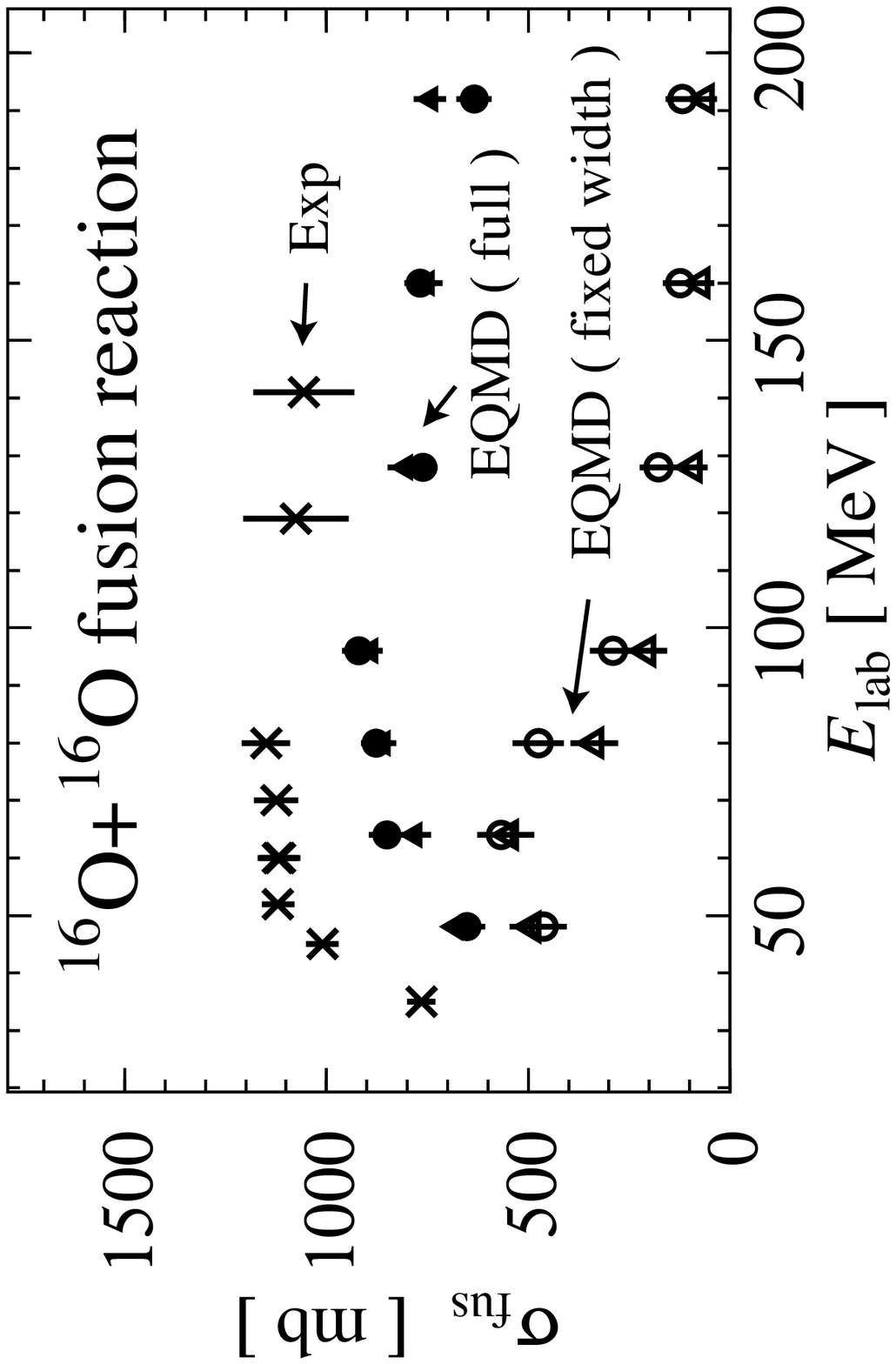}
\vspace{100mm}
\caption {
Fusion cross section for $^{16}{\rm O}+^{16}{\rm O}$ reaction.
Crosses denote experimental data from
Refs.~\protect\cite{refFusionOOa,refFusionOOb}.
The full Circles and full triangles denote our results of
full EQMD calculation with parameter set 1 and 2,respectively.
Open circles and open triangles denote the results of EQMD
calculation with constraint of fixed wave-packet widths.
}
\end{figure}

The fusion cross section is obtained by calculating the fusion
probabilities for impact parameters of 0.0, 3.0, 4.0, 5.0, 5.5
and 6.0 fm by simulating  40 events for each impact parameter.
In Fig.~7 we show the incident-energy dependence of the fusion cross section.
Crosses denote experimental data from Refs.~\protect%
\cite{refFusionOOa,refFusionOOb}
and full circles and triangles denote our calculation
with parameter sets 1 and 2.
There is no significant difference between the results
with two parameter sets.
They reproduce about 85 \% of experimental data but
somewhat underestimate them.
The ground state of $^{16}{\rm O}$ we have used has the binding energy
of 130.8 MeV and the rms radius of 2.5 fm which is about 92 \% of
the experimental value 2.73 fm \cite{refExpCarbonOxigen}.
We guess the small size of $^{16}{\rm O}$ causes this
underestimation of fusion cross section.
Though we can not, at present, reproduce perfectly
the fusion cross section, one should remember that
we get almost zero fusion cross section with the
standard QMD by using the same criterion of fusion event.

We discuss again the effect of dynamical treatment of
wave-packet width by comparing the full calculation with that
of fixed width constraint as in the previous section.
In the fixed width calculation, the initial condition and
interaction is exactly the same as in the full EQMD calculation.
As shown with open circles and triangles in Fig.~7,
the fusion cross sections with fixed width calculation
are close to the full calculation in low-energy region
and are much smaller for higher incident energies.
At very low energies, two nuclei can easily fuse
if they overcome the fusion barrier.
At higher energies, however, the dissipation of the
incident energy into the internal excitation is necessary.
The dynamical change of wave-packet widths offers 
larger degree of freedom for the dissipation than the case of fixed width.
Thus the dynamical treatment of wave-packet width plays much
important roles rather at higher energies where dissipation
of the incident energy is important for the fusion process.

\section{Summary}

In this paper we have discussed how to treat low-energy
reactions in the framework of quantum molecular dynamics (QMD).
We have introduced a phenomenological Pauli potential
into effective interactions and have made the width of
each wave packet as a dynamical variable.
With this extended QMD (EQMD) method, we can well describe the
ground state properties such as binding energies,
density profiles or alpha-clustering
structure in light nuclei.

In the calculation, we have introduced two sets of
interaction parameters.
Set 1 is obtained by adjusting the nuclear potential by keeping the
matter saturation properties which depends both on Pauli potential
and on the nuclear potential.
Set 2 is obtained by the standard parameterization of nuclear potential
and the Pauli potential is searched to give a good agreement to the
systematic trend of binding energies of finite systems.
With the latter parameter set, we can reproduce the binding energies
of nuclei almost completely from very light to very heavy systems.
As for the nucleus-nucleus collisions, however, we did not see any
significant difference between different interaction parameter sets.

The width of Gaussian wave-packets, which are taken
as a constant of arbitrary value in standard QMD,
were determined from the minimum energy condition for the ground state.
We showed that for $^{12}{\rm C}$, the real parts of the widths become
identical for whole 12 nucleons and a clear alpha structure appears
where 3 alphas are located at the vertices of equilateral triangle.
For medium and heavy nuclei the real parts of the width show a
systematic distribution; wider widths in the central region and
narrower ones near the surface, which was first obtained with this work.
Remaining problems to be solved are somewhat higher matter density
near the center and somewhat smaller root mean square radius.

In the calculation of nucleus-nucleus collisions,
we showed that EQMD is able to enhance the
dynamical alpha emission process.
This enhancement was seen in AMD calculation
but not in the standard QMD.

The effect of the dynamical change of the widths of
wave-packets were clearly seen in two points:
One is nucleon emission or disintegration of
clusters in medium energy collisions.
With the dynamical change of Gaussian widths,
nucleons can escape easily from the cluster.
If we freeze the widths, dynamical emission
of nucleons is much hindered.
The other is the fusion reaction where dissipation
of the incident energy become essential.
We showed that the change of widths for fusion of
$^{16}{\rm O}+^{16}{\rm O}$ causes a big enhancement of
the cross section at higher incident energies.
Without this freedom, the EQMD model gives
unphysically small fusion cross section.

The study of EQMD in this paper showed that this is a
promising direction to generalize the QMD simulations.
The systematic calculation of EQMD, however, is now in a beginning stage.
We need further calculations to clarify the basic features of this model.
One of the most interesting item to study with EQMD calculation is
the low energy reactions between heavy nuclei, because no other
many-body model has been successfully applied to it.
The study in this direction is in progress.

\acknowledgements

The authors would thank Dr.~Aldo Bonasera, Prof.~Hisashi Horiuchi and
Dr.~Tomoyuki Maruyama for fruitful discussions.
Support of Institute of Physical and Chemical Research (RIKEN)
for the use of VPP-500 super-computer is also acknowledged.


\begin{references}

\bibitem{refTDHF}For example, J.~W.~Negele, Rev. Mod. Phys. {\bf 54} (1982)
913.

\bibitem{refVUU}For example, G.~F.~Bertsch and S.~Das~Gupta, Phys. Rep.
{\bf 160} (1988) 189; and references therein.

\bibitem{refCascade}For example, Y.~Yariv and Z.~Fraenkel, Phys. Rev. C
{\bf 20} (1979) 2227.

\bibitem{refQMD}J.~Aichelin and H.~St\"ocker, Phys. Lett. B {\bf 176}
(1986) 14.

\bibitem{refQMDrep}J.~Aichelin, Phys. Rep. {\bf 202} (1991) 233;
and references therein.

\bibitem{refFMD}H.~Feldmeier, Nucl. Phys. {\bf A515} (1990) 147.

\bibitem{refAMD}A.~Ono, H.~Horiuchi, T.~Maruyama and A.~Ohnishi,
Prog. Theor. Phys. {\bf 87} (1992) 1185.

\bibitem{refQMDfusion}T.~Maruyama, A.~Ohnishi and H.~Horiuchi,
Phys. Rev. {\bf C42} (1990) 386.

\bibitem{refOhnishi}Similar way to A.~Ohnishi, T.~Maruyama and H.~Horiuchi,
Prog. Theor. Phys. {\bf 87} (1992) 417.

\bibitem{refValta}
P.~Valta, J.~Konopka, A.~Bohnet, J.~Jaenicke, S.~Huber, C.~Hartnack,
G.~Peilert, L.~W.~Neise, J.~Aichelin, H.~St\"ocker and W.~Greiner,
Nucl. Phys. {\bf A538} (1992) 417c.

\bibitem{refPauliA}G.~Peilert, J.~Randrup, H.~Sto\"cker and W.~Greiner,
Phys. Lett. B {\bf 260} (1991) 271.

\bibitem{refPauliB}G.~Peilert, J.~Konopka, H.~St\"ocker,
W.~Greiner, M.~Blann and M.~G.~Mustafa,
Phys. Rev. {\bf C46} (1992) 1457.

\bibitem{refQMDplusStat}T.~Maruyama, A.~Ono, A.~Ohnishi and H.~Horiuchi,
Prog. Theor. Phys. {\bf 87} (1992) 1367.

\bibitem{refExpCarbonOxigen}
I.~Sick and J.~S.~McCarthy, Nucl. Phys. {\bf A150} (1970) 631.

\bibitem{refQMDrel}
K.~Niita, S.~Chiba, T.~Maruyama, T.~Maruyama, H.~Takada,
T.~Fukahori, Y.~Nakahara and A.~Iwamoto, submitted to Phys. Rev. C.

\bibitem{refCascade}
F.~P\"uhlhofer, Nucl. Phys. {\bf A280} (1977) 267.

\bibitem{refFusionOOa}
B.~Fernandez, C.~Gaarde, J.~S.~Larsen, S.~Pontoppidan and F.~Videbaek,
Nucl. Phys. {\bf A306} (1978) 259.

\bibitem{refFusionOOb}
F.~Saint-Laurent, M.~Conjeaud, S.~Harar, J.~M.~Loiseaux,
J.~Menet and J.~B.~Viano,
Nucl. Phys. {\bf A327} (1979) 517.

\end{references}
\end{document}